# Bridging MOOCs, Smart Teaching, and AI-Assisted Learning: A Unified Quantitative Model

Bo Yuan
The University of Queensland
Brisbane, Australia
boyuan@ieee.org

Jiazi Hu
AI U Course
Kunming, P.R. China
jessica@aiucourse.com

***Abstract*—MOOCs, Smart Teaching (ST), and AI-assisted learning support different stages of higher education, yet a unified quantitative basis for analyzing their contributions, course design, and resource allocation remains limited. This study develops a mathematical model integrating MOOC-based pre-class learning, ST-based in-class adaptation, and AI-assisted post-class personalization. Learner mastery is represented as a bounded multidimensional state, with a common exponential learning-response function describing how instructional resources reduce remaining knowledge gaps. The stages differ in their allocation rules: predefined course-content emphasis for MOOCs, feedback-driven class-level adaptation for ST, and individualized gap-based allocation for AI-assisted learning. Simulations across 100 independently generated classes demonstrate stable cumulative progression and quantify stage-wise gains, while budget analysis reveals diminishing returns from additional AI support. For a fixed learner and learning mechanism, varying course-content emphasis shows a strong association between course–learner alignment and post-MOOC mastery. An optimal AI allocation is also derived under a fixed budget: supported components reach a common residual mastery gap, while components below this threshold receive no resources. A controlled comparison yields over 14% greater learning gain than proportional allocation. These analyses make the instructional process quantitatively analyzable and provide a basis for examining course–learner fit and coordinating limited learning resources. The model offers an analytical foundation for instructional decisions, with practical application requiring empirical estimation of learner states and calibration of learning-response parameters.***

## I. INTRODUCTION

Higher education has experienced several waves of digital transformation over the past decade. Massive Open Online Courses (MOOCs) expanded access to structured learning resources and enabled large-scale, self-paced learning [1]. Smart Teaching (ST) subsequently introduced real-time data collection, learning analytics, and interactive technologies into classroom instruction, allowing instructors to respond more directly to emerging learning needs [2]. More recently, advances in generative AI have made scalable, on-demand, and personalized learning possible [3, 4].

These technologies are typically discussed as distinct educational paradigms. MOOCs primarily emphasize scalable access to structured content, but their fixed organization and limited interaction can restrict responsiveness to learners [5]. ST improves instructional visibility by using learning data to guide classroom-level instruction, yet its adaptation is constrained by a fixed amount of teaching time and the need to serve the class as a whole [6]. AI-assisted learning extends adaptation to the individual level by providing personalized feedback and learning support in response to each learner's evolving needs [7]. Viewed separately, these paradigms differ substantially in technological form and pedagogical implementation. Viewed as parts of a continuous learning process, however, they exhibit an important underlying commonality.

This commonality motivates the central argument of this study. At an abstract level, learning in all three settings can be represented as a process in which instructional effort acts on a learner's knowledge gap to produce a change in mastery [8]. What differs is primarily how that instructional effort is allocated. In MOOC-based learning, allocation is largely predetermined by the course structure. In ST, instructional resources can be redistributed according to the aggregate learning state of the class. In AI-assisted learning, allocation can be personalized according to the remaining knowledge gaps of individual students. This suggests that MOOCs, ST, and AI-assisted learning need not be treated merely as separate technologies or loosely connected pedagogical tools. They can instead be modeled as different realizations of a shared learning mechanism with distinct resource-allocation rules.

This perspective addresses an important limitation in existing technology-enhanced learning research. The three paradigms have been extensively studied within their respective contexts, yet their effects are usually evaluated independently. Such separation makes it difficult to analyze how they operate successively within the same learning process and how the contribution of each stage can be distinguished from the final learning outcome. In real educational settings, this problem is further complicated by differences in prior knowledge, learner ability, teacher practice, motivation, course design, and institutional conditions [9]. When several instructional technologies are used together, simple pre/post comparisons provide limited information about where observed gains originate. A unified quantitative representation can also help examine how predefined course emphasis matches learners' needs and whether a given allocation rule makes effective use of a limited learning budget. These questions connect the analysis of learning progression with decisions about course design and instructional resource allocation.

To address this problem, this study develops a unified quantitative model of a three-stage learning process comprising MOOC-based pre-class learning, ST-based in-class adaptation, and AI-assisted post-class personalization. Student knowledge is represented as a bounded multidimensional state, and all three stages are governed by a common exponential learning-response rule. The stages are differentiated through their allocation mechanisms: fixed course-level allocation for MOOCs, class-level adaptive allocation for ST based on aggregate feedback, and individualized allocation for AI-assisted learning according to each learner's remaining mastery gaps. By separating the common learning dynamics from stage-specific allocation rules, the model makes the three instructional modes directly comparable within a single analytical system. Within this formulation, course–learner alignment analysis examines the effect of varying predefined course emphasis while holding the learner and learning mechanism fixed. Optimal AI allocation analysis instead holds the learner state, budget, and response function fixed and determines the resource distribution that maximizes learning gain.

The purpose of this formulation is not to reproduce the full complexity of learning behavior or to predict exact educational outcomes. Rather, it provides a controlled setting in which instructional mechanisms can be isolated and compared, and their respective effects can be quantified. The model therefore focuses on a set of mechanisms that are both pedagogically interpretable and mathematically tractable. More complex features, such as prerequisite relationships [10], dynamic learner strategies [11], feedback across instructional cycles [12], and richer teacher–AI interaction [13], are possible extensions but are beyond the scope of the present study.

The main contributions of this work are threefold. First, it identifies a common learning structure underlying MOOCs, ST, and AI-assisted learning while preserving their distinct pedagogical roles through different resource-allocation rules. Second, it formalizes this structure within a unified mathematical model that enables stage-wise decomposition of learning gains and controlled analysis of instructional allocation. Third, simulations across 100 independently generated classes demonstrate stable learning progression, while individual-level analyses show a strong association between course–learner alignment and learning outcomes and establish the unique globally optimal AI allocation under the model assumptions. A controlled comparison shows over 14% greater learning gain than gap-proportional allocation under the same budget. Together, these results provide a quantitative basis for examining how course emphasis matches learner needs and how limited learning resources can be allocated more effectively.

## II. Learning Mechanisms

MOOCs, ST, and AI-assisted learning differ substantially in technological form and implementation. Rather than reviewing their historical development, this section focuses on the instructional mechanisms that support their integration within a common learning process. From this perspective, the three paradigms are distinguished by how learning resources are allocated across learners and knowledge components.

### A. *MOOCs: Fixed Structured Allocation*

MOOCs provide learners with scalable access to structured educational resources, typically through prerecorded lectures, readings, quizzes, and other asynchronous activities. Their major contribution lies in making organized learning materials widely available while allowing learners to study independently. They are also frequently used as pre-class resources in flipped learning settings, where foundational knowledge acquisition is moved outside the classroom [14, 15].

Despite this flexibility, the relative emphasis placed on different topics, activities, and knowledge components in a MOOC is largely determined in advance through instructional design. Although learners may revisit difficult materials or skip content that they have already mastered [16], this study does not account for such individual navigation choices. Instead, MOOC-based learning is guided by the predefined structure of the course, which places different levels of emphasis across knowledge components. For a given amount of learning effort, the relative distribution of that effort is treated as common to all students and does not change in response to an individual student's knowledge state. Students may nevertheless achieve different learning gains under this shared emphasis as they differ in prior mastery and receptivity to learning activities.

### B. *Smart Teaching: Class-Level Adaptive Allocation*

ST introduces data-driven adaptation into the instructional process. Through learning analytics, classroom interaction data, and formative assessment, instructors can observe the learning state of the class more directly and use this information to adjust instructional emphasis, pacing, and strategies [17, 18]. This represents an important shift from static instructional delivery toward feedback-informed teaching. The key feature of ST is not simply the availability of digital tools, but the use of observed learning evidence to guide instructional decisions. If the class demonstrates insufficient understanding of particular concepts, more classroom time or instructional attention can be directed toward those areas. Conversely, topics that are already well understood may require less additional emphasis.

Such adaptation generally operates under a shared classroom constraint. A teacher reallocates instructional resources for the class as a whole rather than for individual students. ST therefore introduces an allocation mechanism in which instructional resources are no longer fully predetermined, but are adjusted according to the aggregate learning state of the class. From this perspective, the defining mechanism of ST is class-level adaptive allocation based on learning feedback.

### C. *AI-Assisted Learning: Individual-Level Allocation*

Recent generative AI systems extend adaptation beyond the class level by enabling scalable, on-demand support for individual learners. AI-assisted learning environments can provide explanations, answer follow-up questions, generate examples, and offer feedback in response to a learner's immediate needs [19, 20]. Compared with conventional digital resources, these systems can respond dynamically rather than relying entirely on pre-designed instructional content.

This capability changes the granularity of adaptation. Whereas ST uses aggregate evidence to inform instructional decisions at the class level, AI-assisted learning can adapt directly to the knowledge state of an individual learner. Students who have difficulty understanding a particular concept can receive targeted support without requiring corresponding changes for the rest of the class. Instructional effort can therefore be distributed heterogeneously across learners according to their remaining knowledge gaps. The defining mechanism of AI-assisted learning is thus individual-level adaptive allocation. Rather than considering the full spectrum of AI functionalities, such as automated content generation, the present study focuses on its capacity to personalize instructional support according to student-specific knowledge gaps.

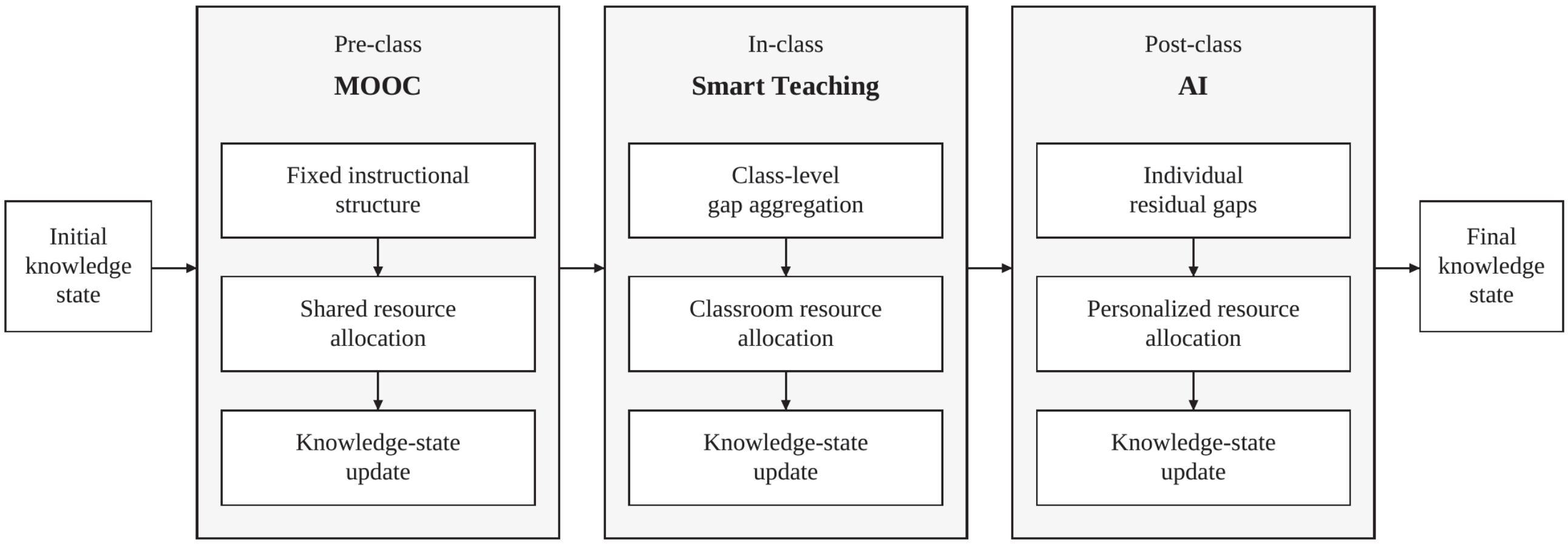

Fig. 1. Overview of the unified instructional framework, showing knowledge-state progression through pre-class MOOC learning, class-level adaptation in Smart Teaching, and personalized post-class AI support.

### D. *Shared Learning Mechanism*

Although these approaches differ in implementation, they can be interpreted through a common logic. In each case, instructional resources act on the learner's current knowledge state to reduce remaining mastery gaps, while the primary distinction lies in how those resources are distributed.

MOOC-based learning generally follows a fixed allocation determined by a predefined course-content emphasis vector. ST introduces feedback-driven redistribution at the class level, whereas AI-assisted learning increases the granularity of adaptation by allocating support at the individual level. This perspective suggests that they need not be modeled through fundamentally different learning processes. Instead, they can be represented as distinct realizations of a shared learning-response mechanism with different allocation rules. This structure provides the basis for a unified mathematical formulation while preserving their distinct instructional roles.

The next section organizes these mechanisms into a continuous three-stage learning process. This pedagogical structure provides the basis for the quantitative formalization developed in Section IV.

## III. INTEGRATED LEARNING PROCESS

### A. *A Natural Three-Stage Sequence*

In higher-education settings, MOOCs, ST, and AI-assisted learning can naturally appear in a sequential relationship. Students may first use structured MOOC resources for pre-class preparation. Instructors then use ST during class to respond to the aggregate learning state of the group, and AI-assisted learning can provide individualized support after class based on remaining learning needs. This sequence also reflects a progressive increase in the granularity of instructional adaptation, moving from predefined course-level structure to class-level adjustment and finally to individual-level support.

Fig. 1 summarizes this sequence, showing the progression of knowledge states and the resource allocation mechanism associated with each stage. This sequence reflects a common instructional pattern rather than an artificial combination of three technologies. MOOC-based learning is well suited to structured preparation because course materials can be accessed asynchronously before class. ST becomes relevant during classroom instruction, where the teacher can use learning evidence to adjust instructional emphasis for the class as a whole. After class, AI-assisted learning can further extend support to the individual level by responding to student-specific questions, misunderstandings, and remaining knowledge gaps.

From the perspective of information flow, the process moves from individual learning states to class-level aggregation and then back to individual-level personalization. Students first develop heterogeneous post-MOOC knowledge states. These states can be summarized to provide class-level feedback for ST, after which each student's updated state can again serve as the basis for individualized AI support. This creates a coherent instructional sequence in which each stage operates at a different level while contributing to the same learning process.

### B. *Motivation for Quantitative Analysis*

Once these stages are considered as parts of a continuous learning process, an important analytical challenge arises. The final learning outcome reflects the cumulative effect of multiple instructional mechanisms. In practice, it is therefore difficult to determine how much improvement originates from each stage of the learning process.

Quantitative models have been developed to characterize learning and guide instruction. Bayesian knowledge tracing incorporates student-specific parameters to model evolving mastery from performance data [21], while deep knowledge tracing learns representations of student knowledge from interaction sequences to predict subsequent performance [22]. Decision-theoretic approaches further use learner models to plan teaching actions under uncertainty [23]. These studies provide foundations for learner modeling and adaptive instruction.

The present study addresses a complementary question: how different resource-allocation mechanisms jointly shape learning progression across successive instructional stages. Each stage starts from the learning state produced by the preceding stage, so the effect of an intervention depends both on its allocation rule and on the mastery gaps that remain. The proposed quantitative model makes these dependencies explicit by representing student mastery as a state that evolves through a common learning-response function. Within this shared structure, MOOCs, ST, and AI-assisted learning follow distinct rules for fixed course-level allocation, class-level adaptation, and individualized support. This allows stage-wise gains to be calculated, allocation mechanisms to be compared under controlled conditions, and the sensitivity of outcomes to learning budgets to be examined. The same formulation also supports analysis of how course emphasis aligns with learner needs and how individualized resources can be optimally distributed under a fixed budget. Its value lies in making the instructional process quantitatively analyzable, providing a basis for evaluating course–learner fit and examining resource-allocation decisions alongside stage-wise learning contributions.

## IV. Quantitative Formalization

### *A. Modeling Rationale and State Representation*

To enable quantitative analysis of the learning process, we introduce a simplified mathematical model that represents how students' knowledge states evolve under different forms of instructional support. The purpose is not to reproduce every cognitive, behavioral, or contextual factor involved in real learning. Rather, it aims to retain several essential characteristics of educational practice while removing factors that would otherwise make controlled analysis difficult.

In real educational settings, observed learning outcomes are jointly influenced by prior knowledge, learner ability, teacher practice, course design, classroom composition, motivation, and many other factors [24]. A controlled mathematical model provides a complementary means of studying these mechanisms by specifying how knowledge states evolve across stages, how instructional resources are allocated, and how students respond to those resources [25].

We consider a class of $N$ students studying a course composed of $d$ knowledge components. The knowledge state of student $i$ is represented by

$$\mathbf{k}^i = [k_1^i, k_2^i, \ldots, k_d^i] \in [0,1]^d, \quad i = 1, \ldots, N,$$

where $k_j^i$ denotes the mastery level of student $i$ for knowledge component $j$. A value of 0 represents no mastery, whereas a value of 1 represents full mastery. The initial state of student $i$ is denoted by $\mathbf{k}_0^i$. After the MOOC-based learning stage, the corresponding state is denoted by $\mathbf{k}_{\mathrm{M}}^i$. Following the ST stage, the state becomes $\mathbf{k}_{\mathrm{S}}^i$ , and after AI-assisted learning, the final state is denoted by $\mathbf{k}_{\mathrm{A}}^i$.

To capture differences in learning responsiveness, each student is assigned a receptivity parameter $\alpha_i \in (0, 1]$, which characterizes the efficiency with which instructional exposure is converted into knowledge gain. In practice, it may reflect many underlying factors, including prior experience, cognitive ability, motivation, study habits, and engagement [26]. Modeling these factors individually would require a complex learner model and introduce parameters that are difficult to identify reliably.

Each stage is associated with a total learning budget $\tau_{\mathrm{X}}$, where $\mathrm{X} \in \{\mathrm{M}, \mathrm{S}, \mathrm{A}\}$ indexes the MOOC, ST, and AI-assisted learning stages, respectively. To keep the average learning budget per component constant as $d$ varies, the total budget is linearly scaled with $d$. Learning gains are assumed to depend on the remaining mastery gap and, for knowledge component $j$, the quantity 1-$k_j^i$ represents the opportunity for improvement. This gap-based formulation is consistent with established evidence of diminishing gains from continued practice [27] and bounds the potential gain for each component, without requiring a detailed cognitive model of learning.

Knowledge components are further assumed to be separable dimensions. Specifically, improvement in one component does not directly alter the mastery of another component. Under this assumption, the learning dynamics of component $j$ depend only on its own current mastery and the instructional support directed toward it. Although cross-concept transfer may exist in real curricula, component-wise representations of learner mastery are widely used in student modeling and knowledge tracing [28]. The separability assumption adopted here therefore provides a simplified approximation that preserves analytical tractability. More complex dependency structures can be incorporated in future extensions without altering the modeling principle.

In addition to individual learning states, the model introduces a class-level learning state and the aggregate mastery of the class is represented by $\overline{\mathbf{k}} = \frac{1}{N}\sum_{i=1}^{N} \mathbf{k}^i$. In particular, after the MOOC stage, $\overline{\mathbf{k}}_{\mathrm{M}} = \frac{1}{N}\sum_{i=1}^{N} \mathbf{k}_{\mathrm{M}}^i$ represents the mastery profile of the class across the $d$ knowledge components. This distinction is important as different instructional mechanisms operate at different levels of granularity. Classroom teaching typically responds to aggregated evidence about the learning status of the class, whereas AI-assisted learning can provide support at the level of individual students.

The learning process is therefore modeled as a sequence of state transformations

$$\{\mathbf{k}_0^i\}_{i=1}^N \to \{\mathbf{k}_{\mathrm{M}}^i\}_{i=1}^N \to \overline{\mathbf{k}}_{\mathrm{M}} \to \{\mathbf{k}_{\mathrm{S}}^i\}_{i=1}^N \to \{\mathbf{k}_{\mathrm{A}}^i\}_{i=1}^N.$$

For student $i$ and knowledge component $j$, the generic update is defined as

$$k_{j,new}^i = k_j^i + [1 - \exp(-\alpha_i \tau w_j)](1 - k_j^i), \qquad (1)$$

where $\tau \geq 0$ is the total learning budget of the current stage and $w_j \geq 0$ denotes the proportion of the budget allocated to component $j$, with $\sum_{j=1}^{d} w_j = 1$ The term $1 - \exp(-\alpha_i \tau w_j)$ represents the fraction of the remaining mastery gap that is closed under the given instructional intensity.

Since $\alpha_i \tau w_j \geq 0$, we have

$$0 \leq [1 - \exp(-\alpha_i \tau w_j)](1 - k_j^i) \leq 1 - k_j^i.$$

This update guarantees non-negative learning gains and prevents the mastery state from exceeding its upper bound without requiring any additional clipping operation.

*B. MOOC-Based Structured Self-Learning*

The first stage models pre-class self-learning through MOOCs, which is a consequential part of flipped learning and has been found to strongly influence its overall success [29]. We assume that each student is assigned the same learning budget and follows the instructional structure predefined by the course. The MOOC is therefore modeled as a common and non-adaptive learning environment: all students are exposed to the same set of knowledge components with the same relative emphasis, although their resulting learning gains may differ because of variations in prior mastery and learning receptivity.

Let $\mathbf{w}_{\mathrm{C}} = [w_{\mathrm{C},1}, w_{\mathrm{C},2}, \dots, w_{\mathrm{C},d}] \in [0,1]^d$ denote the course-content emphasis vector, where $w_{\mathrm{C},j}$ represents the relative proportion of the learning budget assigned to component $j$. The vector is normalized as $\sum_{j=1}^{d} w_{\mathrm{C},j} = 1$, which reflects the distribution of instructional emphasis across knowledge components, such as the relative amount of video content, reading material, or practice associated with each component. For analytical clarity, all students are assumed to follow this common structure during the MOOC stage.

Given the initial state $\mathbf{k}_0^i$ of student $i$, the post-MOOC knowledge state is

$$\mathbf{k}_{\mathrm{M}}^i = \mathbf{k}_0^i + [\mathbf{1} - \exp(-\alpha_i \tau_{\mathrm{M}} \mathbf{w}_{\mathrm{C}})] \odot (\mathbf{1} - \mathbf{k}_0^i), \quad (2)$$

where $\odot$ denotes element-wise multiplication and $\mathbf{1}$ denotes the all-ones vector in $\mathbb{R}^d$. In Eq. (2), the exponential function is also applied element-wise.

For knowledge component $j$, the update can be written as

$$k_{\mathrm{M},j}^i = k_{0,j}^i + [1 - \exp(-\alpha_i \tau_{\mathrm{M}} w_{\mathrm{C},j})](1 - k_{0,j}^i). \quad (3)$$

Learning gains increase with both the total pre-class learning budget $\tau_{\mathrm{M}}$ and the allocation weight $w_{\mathrm{C},j}$, while the exponential response introduces diminishing returns as instructional intensity increases. After the MOOC stage, the class-level state is computed as $\overline{\mathbf{k}}_{\mathrm{M}} = \frac{1}{N}\sum_{i=1}^{N} \mathbf{k}_{\mathrm{M}}^i$, which provides the learning feedback used in the ST stage.

*C. Smart Teaching As Class-Level Adaptive Instruction*

The second stage models in-class adaptive instruction. Unlike the MOOC stage, ST uses feedback from the class to adjust how classroom resources are distributed across knowledge components at the class level [30]. Let the post-MOOC class state vector be

$$\overline{\mathbf{k}}_{\mathrm{M}} = [\overline{k}_{\mathrm{M},1}, \overline{k}_{\mathrm{M},2}, \dots, \overline{k}_{\mathrm{M},d}],$$

where $\overline{k}_{\mathrm{M},j}$ represents the average mastery of knowledge component $j$ across the class. Components with lower class-level mastery indicate greater collective learning needs, while the attention that they actually receive also depends on the predefined course-content emphasis. We define the ST allocation vector as

$$\mathbf{w}_{\mathrm{S}} = [w_{\mathrm{S},1}, w_{\mathrm{S},2}, \dots, w_{\mathrm{S,d}}], \quad \text{with } \sum_{j=1}^{d} w_{\mathrm{S,j}} = 1,$$

where $w_{\mathrm{S,j}}$ denotes the proportion of the fixed in-class instructional time allocated to knowledge component $j$.

To reflect adaptive resource redistribution, we define

$$\mathbf{w}_{\mathrm{S}} = \frac{\mathbf{w}_{\mathrm{C}} \odot (\mathbf{1} - \overline{\mathbf{k}}_{\mathrm{M}})}{\|\mathbf{w}_{\mathrm{C}} \odot (\mathbf{1} - \overline{\mathbf{k}}_{\mathrm{M}})\|_1}. \quad (4)$$

Here, $\|\cdot\|_1$ denotes the $\ell_1$-norm and the resource allocation is determined jointly by $\mathbf{w}_{\mathrm{C}}$ and the observed class-level gap $\mathbf{1} - \overline{\mathbf{k}}_{\mathrm{M}}$. If the denominator is zero, the corresponding allocation weights are set to $1/d$. Since no mastery gap remains, the knowledge state is unchanged.

All students are exposed to the same classroom allocation $\mathbf{w}_{\mathrm{S}}$, reflecting the practical constraint that a teacher typically adapts instruction for the class as a whole rather than providing a fully individualized teaching sequence for every student. However, the learning gains remain student-specific because each student enters the classroom with a different post-MOOC state and receptivity parameter.

For student $i$, the post-ST knowledge state is modeled as

$$\mathbf{k}_{\mathrm{S}}^i = \mathbf{k}_{\mathrm{M}}^i + [\mathbf{1} - \exp(-\alpha_i \tau_{\mathrm{S}} \mathbf{w}_{\mathrm{S}})] \odot (\mathbf{1} - \mathbf{k}_{\mathrm{M}}^i). \quad (5)$$

Equivalently, for knowledge component $j$,

$$k_{\mathrm{S},j}^i = k_{\mathrm{M},j}^i + [1 - \exp(-\alpha_i \tau_{\mathrm{S}} w_{\mathrm{S},j})](1 - k_{\mathrm{M},j}^i). \quad (6)$$

This stage differs fundamentally from MOOC learning in its decision mechanism. In ST, the instructional resource is reallocated in response to the observed class-level learning state. The adaptation is therefore feedback-driven: aggregate learning evidence is used to identify where the class requires greater instructional attention, while the resulting allocation remains common to all students. ST is therefore modeled as a class-level adaptive allocation mechanism operating under a shared classroom resource constraint.

*D. AI-Assisted Individualized Learning*

The third stage models post-class learning supported by an AI teaching assistant. While ST adapts instructional resources to class-level needs, the AI assistant tailors support to each student's remaining knowledge gaps [31]. Accordingly, the post-class learning budget is allocated separately for each student across knowledge components.

For student $i$, we define an AI allocation vector

$$\mathbf{w}_{\mathrm{A}}^i = [w_{\mathrm{A},1}^i, w_{\mathrm{A},2}^i, \dots, w_{\mathrm{A},d}^i], \quad \text{with } \sum_{j=1}^{d} w_{\mathrm{A},j}^i = 1.$$

Unlike the ST allocation, which remains influenced by the original course-content emphasis vector, the AI allocation is determined solely by the student's post-ST knowledge state and we therefore define

$$\mathbf{w}_{\mathrm{A}}^i = \frac{\mathbf{1} - \mathbf{k}_{\mathrm{S}}^i}{\|\mathbf{1} - \mathbf{k}_{\mathrm{S}}^i\|_1}. \quad (7)$$

This formulation assigns more AI-assisted learning effort to knowledge components with larger remaining mastery gaps. A student who has nearly mastered a component receives little additional support for that component, whereas weaker components receive greater attention. Since the allocation is derived independently for each student, different students generally obtain different AI learning profiles. If all components are fully mastered, the weights are set to $1/d$ and the knowledge state remains unchanged because no mastery gap remains.

Given the total AI learning budget $\tau_{\mathrm{A}}$, the post-AI state of student $i$ is modeled as

$$\mathbf{k}_{\mathrm{A}}^{i} = \mathbf{k}_{\mathrm{S}}^{i} + \left[\mathbf{1} - \exp\left(-\alpha_i \tau_{\mathrm{A}} \mathbf{w}_{\mathrm{A}}^{i}\right)\right] \odot \left(\mathbf{1} - \mathbf{k}_{\mathrm{S}}^{i}\right). \tag{8}$$

For knowledge component $j$,

$$k_{\mathrm{A},j}^{i} = k_{\mathrm{S},j}^{i} + \left[1 - \exp\left(-\alpha_i \tau_{\mathrm{A}} \mathrm{w}_{\mathrm{A},j}^{i}\right)\right]\left(1 - k_{\mathrm{S},j}^{i}\right). \tag{9}$$

The budget $\tau_{\mathrm{A}}$ determines the total amount of AI learning effort, while $\mathbf{w}_{\mathrm{A}}^{i}$ determines how that effort is personalized across components. This allows support to be redirected more precisely toward residual individual needs after class-level instruction has already addressed common learning gaps.

### *E. Integrated Quantitative Formulation*

The proposed formulation represents the learning process as a sequence of functionally distinct instructional stages governed by a common bounded learning-response mechanism. Their differences are captured through stage-specific allocation rules, allowing the effects of structured preparation, class-level adaptation, and individual personalization to be represented within the same mathematical system. This unified formulation enables the contribution of each stage to be separated from the cumulative learning outcome and allows the effects of different allocation mechanisms to be examined under controlled assumptions. Because both instructional allocation and learner state are represented explicitly, the model can also be used to examine how their alignment affects learning outcomes. The following simulation examines stage-wise progression, adaptive resource allocation, budget sensitivity, robustness across heterogeneous learner populations, and course–learner alignment at the individual level.

## V. Simulation Study

The simulation study first examines the complete three-stage learning process at the class level and then uses the same formulation to analyze course–learner alignment for a fixed individual learner.

### *A. Simulation Setup*

A class of $N$=100 students and $d$=10 components was considered. Initial mastery was generated by combining three sources of variation. First, each student was assigned an overall prior knowledge level $\mu_i \sim \mathrm{Beta}(2,2)$, representing differences in background. Second, a common mastery profile was used to represent differences in prior mastery across components

$\mathbf{b}$=[0.65, 0.38, 0.57, 0.32, 0.60, 0.42, 0.30, 0.68, 0.50, 0.26].

Finally, a random term $\epsilon_{ij} \sim \mathcal{N}(0, 0.08^2)$ introduced variation specific to each student–component pair. The initial mastery of student $i$ on component $j$ was calculated as

$$k_{0,j}^{i} = \mathrm{clip}\left(0.55\mu_i + 0.45 b_j + \epsilon_{ij}, 0, 1\right),$$

where $\mathrm{clip}(\cdot, 0, 1)$ constrains the result to $[0,1]$. Thus, each student's initial mastery profile reflects both the personal prior knowledge and the shared pattern across components, with additional individual variation.

Each student was also assigned a receptivity parameter $\alpha_i \sim \mathrm{U}(0.6, 1.0)$, which governs their response to subsequent learning activities. The predefined instructional emphasis across knowledge components was specified by the course-content emphasis vector

$\mathbf{w}_{\mathrm{C}}$=[0.08, 0.12, 0.07, 0.15, 0.09, 0.10, 0.13, 0.06, 0.11, 0.09],

where each entry represents the proportion of MOOC learning resources allocated to the corresponding component.

All stages were assigned the same per-component learning budget $\rho_{\mathrm{M}} = \rho_{\mathrm{S}} = \rho_{\mathrm{A}} = 0.5$. The total budget for each stage was calculated as $\tau_{\mathrm{X}} = d\rho_{\mathrm{X}}$, resulting in $\tau_{\mathrm{M}} = \tau_{\mathrm{S}} = \tau_{\mathrm{A}} = 5$. Using equal budgets provides a controlled comparison in which differences among the three stages arise primarily from their allocation mechanisms rather than from differences in total learning effort. A fixed random seed was used to generate a simulated class for experiments shown in Figs. 2–4. These settings are used for the class-level experiments in Sections V, while Section V-F uses a fixed representative learner to isolate the effect of course–learner alignment.

### *B. Stage-Wise Learning Progression*

Fig. 2 shows the distribution of student overall mastery across the four successive states, calculated as the average mastery across all knowledge components $k^i = \frac{1}{d}\sum_{j=1}^{d} k_j^i$. For the simulated class, the class mean mastery increases from 0.498 initially to 0.666 after MOOC learning, 0.775 after ST, and 0.851 after AI-assisted learning.

The results illustrate the cumulative nature of the proposed three-stage process. MOOC learning provides the first substantial improvement through structured pre-class exposure. ST further increases mastery by reallocating classroom resources according to the learning state of the class. AI-assisted learning then provides an additional improvement through individualized post-class allocation based on each student's remaining knowledge gaps. The decreasing absolute gains across successive stages should not be interpreted as indicating decreasing instructional value. As each stage begins from a progressively higher mastery level, the remaining learnable gap becomes smaller, which naturally leads to diminishing absolute gains as mastery approaches its upper limit.

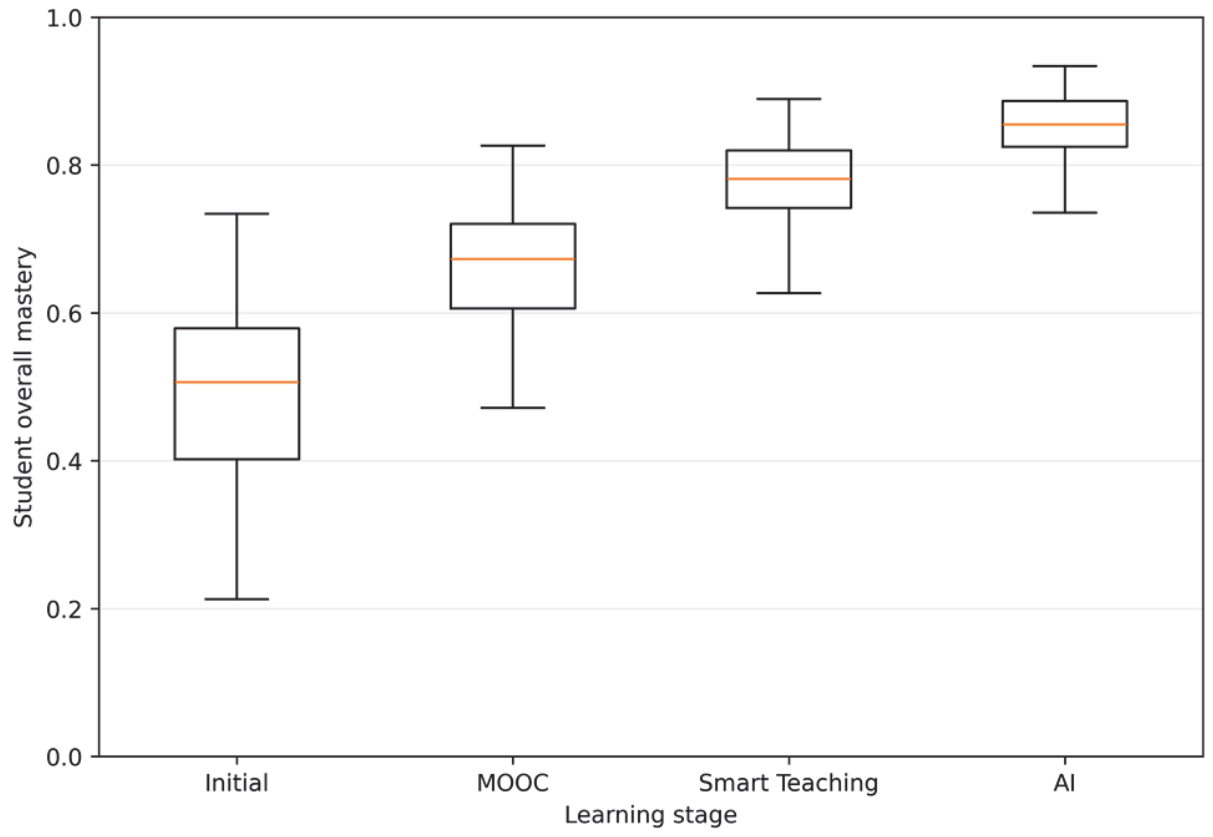

Fig. 2. Distribution of student overall mastery across the four learning stages for a simulated class of 100 students.

## C. Class-Level Adaptation in Smart Teaching

Fig. 3 examines the internal allocation mechanism of the ST stage. The hollow bars show the post-MOOC class-level knowledge gaps, while the line with markers shows the corresponding ST allocation weights. The figure shows that ST allocation is not determined by knowledge gaps alone. Instead, it jointly reflects the remaining class-level learning needs and the predefined course-content emphasis.

For example, knowledge components 4 and 5 have similar post-MOOC knowledge gaps, approximately 0.305 and 0.307, respectively, but receive markedly different allocation weights of approximately 0.137 and 0.083. The higher allocation to component 4 reflects its greater predefined course-content emphasis despite the comparable mastery gaps. This result illustrates a key property of ST: class-level adaptation balances observed learning needs with instructional priorities established in the course design. It therefore reallocates resources in response to class-level feedback without discarding the underlying instructional design.

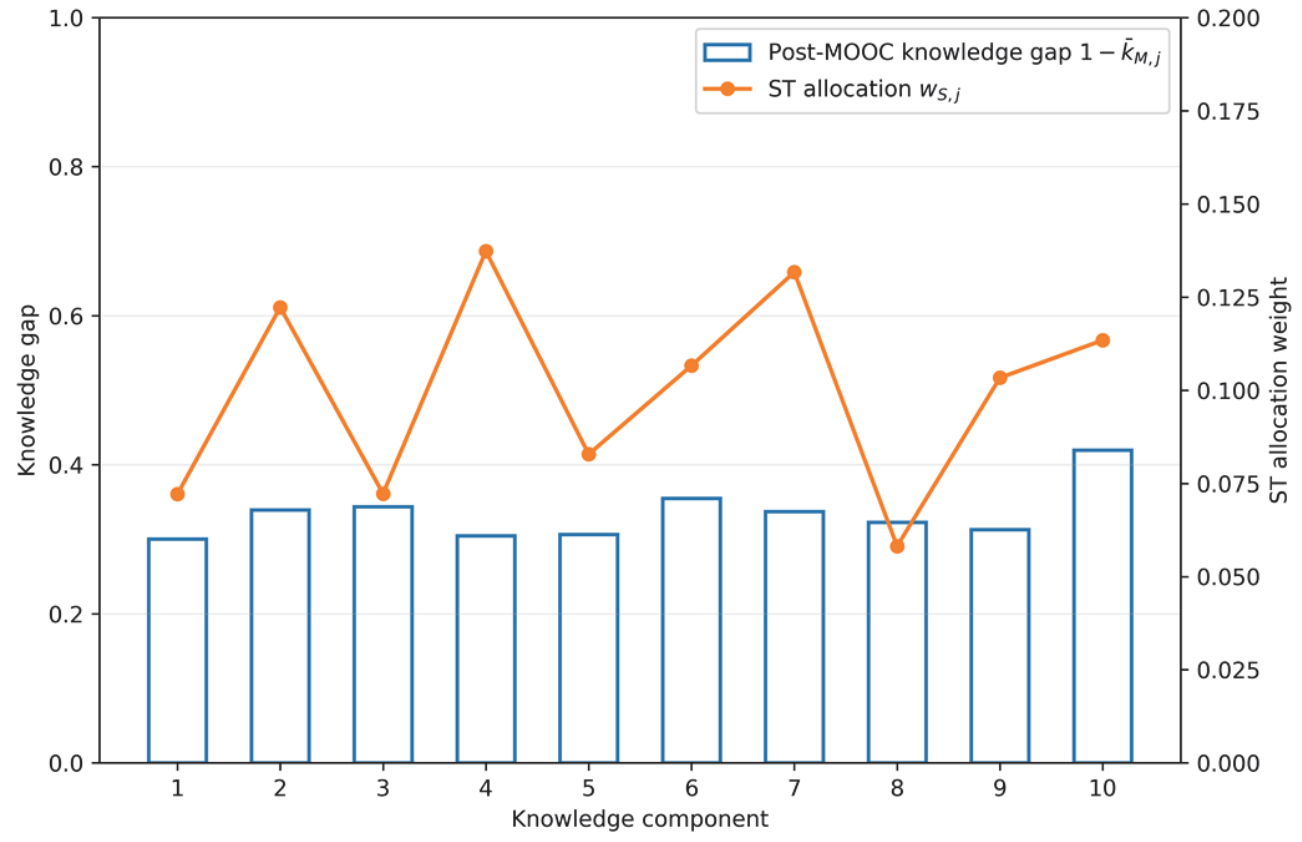

Fig. 3. Post-MOOC class-level knowledge gaps and the corresponding ST allocation weights, which jointly reflect learning needs and course priorities.

## D. Effect of Learning Budget

Fig. 4 analyzes the sensitivity of the final learning outcome to the amount of AI-assisted post-class learning by varying the per-component learning budget $\rho_A$ from 0 to 1, with $\tau_A = d\rho_A$. When $\rho_A = 0$, no AI-assisted learning is introduced, and the class mean mastery remains at 0.775, which is consistent with post-ST stage. As $\rho_A$ increases, final mastery rises continuously. At the baseline setting $\rho_A$ =0.5, the mean mastery reaches approximately 0.851, while increasing $\rho_A$ to 1.0 raises it further to approximately 0.900.

The marginal improvement becomes smaller as the AI learning budget increases. This behavior follows directly from the bounded exponential learning response introduced in Section IV. As students close a larger proportion of their remaining knowledge gaps, additional learning effort has less remaining room to produce further gains. The result is pedagogically useful because it distinguishes between the availability of additional AI learning support and its marginal benefit. Increasing AI-assisted learning opportunities can continue to improve mastery, but the model does not imply proportional returns from additional learning time.

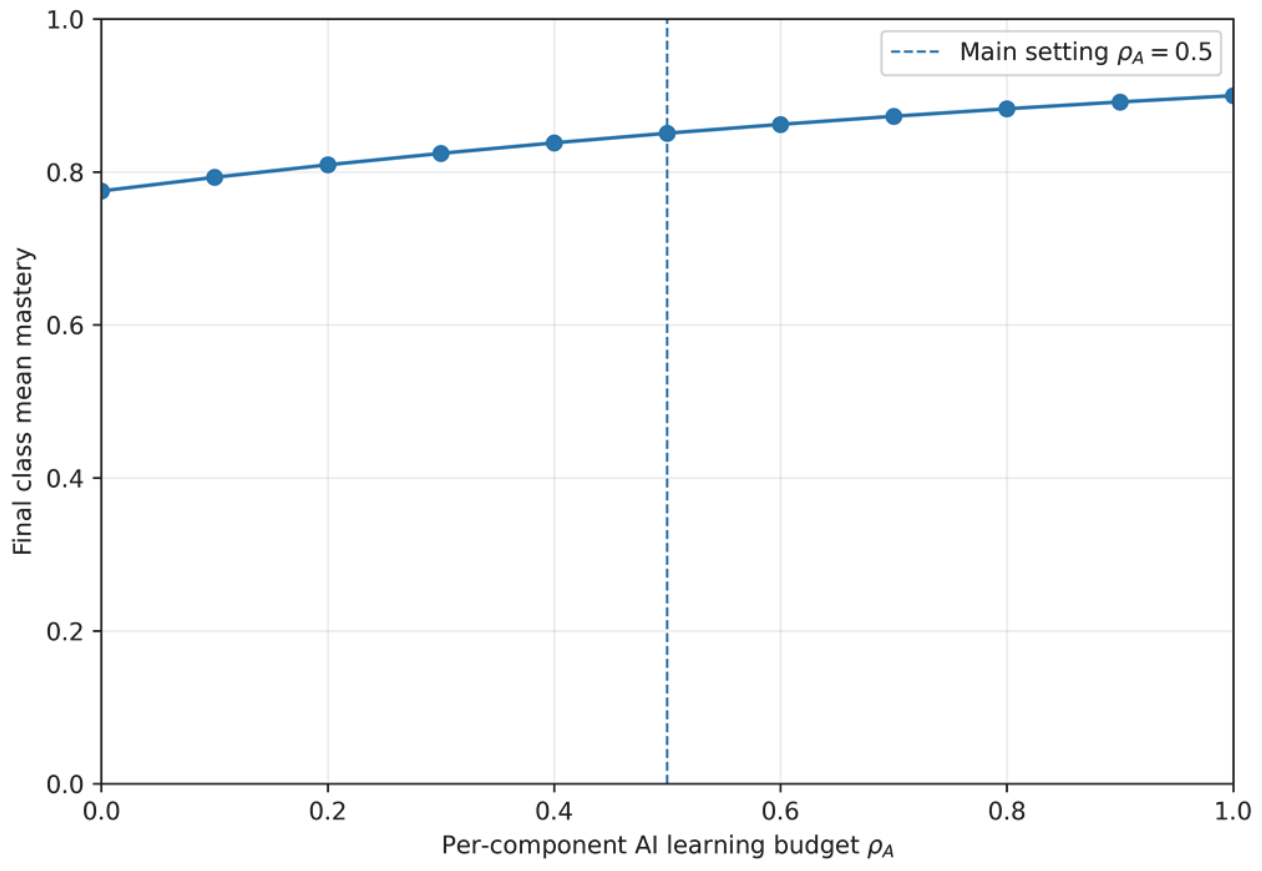

Fig. 4. Sensitivity of final class mean mastery to the per-component AI learning budget, showing diminishing returns as the budget increases.

## E. Robustness Analysis

The learning process was repeated for 100 independently generated classes. Table I reports the class mean mastery at the initial, post-MOOC, post-ST, and post-AI stages. For each stage, the mean, standard deviation, and 95% confidence interval were calculated. The class mean mastery increases consistently across the four states, from 0.485 at the initial state to 0.655 after MOOC learning, 0.767 after ST, and 0.844 after AI-assisted learning. The corresponding mean gains are approximately 0.170, 0.112, and 0.077 for the three instructional stages, respectively. These stage-wise gains provide a quantitative decomposition of the cumulative learning progression, allowing the additional contribution associated with each successive stage to be identified explicitly rather than inferred only from the difference between the initial and final states.

The relatively small standard deviations and narrow confidence intervals indicate that this progression remains stable across classes. The smaller absolute gains in later stages should not be interpreted as indicating weaker instructional value. Each stage operates on the mastery state produced by the preceding stage, leaving progressively smaller knowledge gaps to be addressed. Under the bounded exponential learning response, diminishing absolute gains are therefore an expected consequence of the sequential accumulation of mastery.

TABLE I. STAGE-WISE CLASS MEAN MASTERY

| Stage | Mean | SD | 95% CI |
| --- | --- | --- | --- |
| Initial | 0.485 | 0.0118 | [0.482, 0.487] |
| MOOC | 0.655 | 0.00821 | [0.654, 0.657] |
| ST | 0.767 | 0.00609 | [0.766, 0.768] |
| AI | 0.844 | 0.00464 | [0.843, 0.845] |

## F. Course–Learner Alignment

Adaptive learning emphasizes the importance of matching instructional content and learning pathways to learners' prior knowledge and current needs [32]. Within the present model, this relationship can be examined directly through the alignment between the predefined MOOC course-content emphasis and an individual learner's mastery gaps. The preceding experiments focus primarily on learning progression and resource allocation at the class level. Here, we consider a fixed individual learner and isolate the MOOC stage to examine how the effectiveness of an instructional structure depends on its correspondence with the learner's current knowledge state.

We consider a single learner and omit the learner index for clarity. Given the initial knowledge state $\mathbf{k}_0$, the remaining learning need is represented by the residual-gap vector $\mathbf{g}_0 = \mathbf{1} - \mathbf{k}_0$. The MOOC stage allocates its learning budget according to the course-content emphasis vector $\mathbf{w}_\mathrm{C}$, as defined in Section IV-B. We quantify the correspondence between the course structure and the learner's needs using cosine similarity:

$$A = \frac{\langle \mathbf{w}_\mathrm{C}, \mathbf{g}_0 \rangle}{\|\mathbf{w}_\mathrm{C}\|_2 \|\mathbf{g}_0\|_2}, \quad (10)$$

where $A$ denotes the course–learner alignment. A larger value indicates that greater instructional emphasis is placed on components for which the learner has larger mastery gaps.

To isolate this relationship, a fixed representative learner was used with

$$\mathbf{k}_0 = [0.32, 0.58, 0.45, 0.21, 0.64, 0.39, 0.53, 0.26, 0.48, 0.61],$$

giving an initial mean mastery of 0.447. The corresponding residual-gap vector is

$$\mathbf{g}_0 = [0.68, 0.42, 0.55, 0.79, 0.36, 0.61, 0.47, 0.74, 0.52, 0.39].$$

The learner receptivity was fixed at $\alpha$=0.8, and the MOOC learning budget was set to the same baseline value used in the preceding simulations $\tau_\mathrm{M} = d\rho_\mathrm{M} = 5$. A large set of random course-content emphasis vectors was generated, covering a broad range of alignment levels. For each $\mathbf{w}_\mathrm{C}$, the learner's post-MOOC knowledge state was calculated using the same update rule as Eq. (2). The learning outcome was measured by the mean post-MOOC mastery $\bar{k}_\mathrm{M} = \frac{1}{d}\sum_{j=1}^{d} k_{M,j}$.

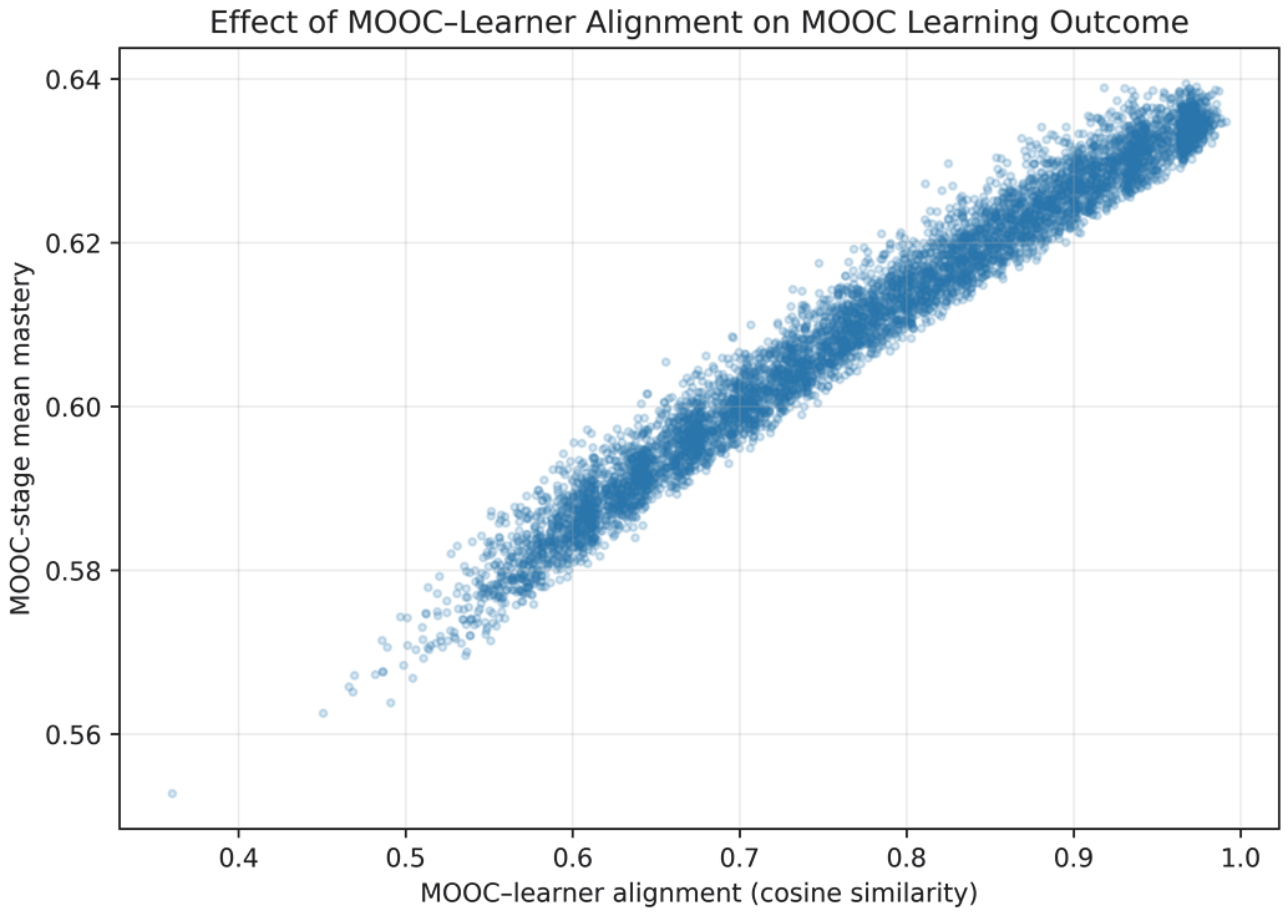

Fig. 5. Relationship between MOOC–learner alignment and post-MOOC mean mastery for a fixed learner.

Fig. 5 shows the relationship between course–learner alignment and post-MOOC mean mastery. The courses produced alignment values ranging from approximately 0.36 to 0.99. Over this range, post-MOOC mean mastery varied from approximately 0.553 to 0.639, corresponding to a spread of about 0.086. A strong positive association is observed between alignment and learning outcome, with Pearson correlation r = 0.986. The result indicates that MOOC effectiveness depends strongly on how well the predefined instructional emphasis corresponds to the learner's remaining mastery gaps. Higher alignment generally leads to greater learning gains because more of the available instructional effort is directed toward components with larger opportunities for improvement.

Different course-content vectors can have similar alignment while distributing resources differently across individual knowledge components. Alignment therefore provides a compact measure of course–learner fit rather than a complete description of instructional allocation. The approximately linear relationship in Fig. 5 is also consistent with the MOOC update rule. For knowledge component $j$, the learning gain is

$$\Delta k_j = [1 - \exp(-\alpha \tau_\mathrm{M} w_{\mathrm{C},j})] g_{0,j}, \quad (11)$$

where $g_{0,j}$ is the $j$-th component of the residual-gap vector $\mathbf{g}_0$. Accordingly, the mean MOOC gain can be written as

$$\Delta \bar{k}_\mathrm{M} = \frac{1}{d}\sum_{j=1}^{d} g_{0,j} [1 - \exp(-\alpha \tau_\mathrm{M} w_{\mathrm{C},j})]. \quad (12)$$

For moderate instructional weights, the exponential term can be approximated using the first-order Taylor approximation,

$$1 - \exp(-\alpha \tau_\mathrm{M} w_{\mathrm{C},j}) \approx \alpha \tau_\mathrm{M} w_{\mathrm{C},j},$$

yielding

$$\Delta \bar{k}_\mathrm{M} \approx \frac{\alpha \tau_\mathrm{M}}{d} \langle \mathbf{w}_\mathrm{C}, \mathbf{g}_0 \rangle. \quad (13)$$

This approximation relates the mean learning gain to the inner product between course-content emphasis and the learner's residual gaps. Since $\langle \mathbf{w}_\mathrm{C}, \mathbf{g}_0 \rangle = A \|\mathbf{w}_\mathrm{C}\|_2 \|\mathbf{g}_0\|_2$, cosine alignment captures one aspect of this relationship, while the norm of the course-content vector varies across courses. The strong positive association observed in Fig. 5 characterizes the sampled course designs rather than an exact proportional relationship determined by alignment alone. The nonlinear exponential response and differences in resource distribution contribute to the scatter around the dominant trend. Taken together, this individual-level analysis complements the class-level experiments by demonstrating that the same quantitative formulation can be used not only to analyze stage-wise learning effects, but also to examine how a predefined course structure interacts with the learning needs of a particular student.

## VI. Optimal Allocation Analysis

The preceding simulations use the gap-proportional AI allocation defined in Eq. (7). This rule provides an intuitive and simple way to personalize learning support: components with larger remaining mastery gaps receive more resources. However, allocating resources in proportion to learning needs does not necessarily maximize the overall learning gain under a fixed budget. This section examines this distinction and derives an optimal allocation under the same learning-response model.

### *A. From Mastery Gaps to Marginal Gains*

Proportional allocation is reasonable as a direct response to individual learning needs, but it does not explicitly account for the marginal gain produced by each additional unit of resource. Under the exponential response, this marginal gain is proportional to the remaining mastery gap and decreases as resources are invested. Allocation should therefore consider not only the initial gaps but also how they change with resource investment. For instance, if two components receiving resources have different remaining gaps, transferring a small amount of resource from the component with the smaller gap to the one with the larger gap increases the overall learning gain: the additional gain in the latter exceeds the gain lost in the former. As a result, at the optimum, all components receiving resources must have the same final marginal gain and, under the common receptivity parameter, the same remaining mastery gap.

With a limited budget, some components may already have gaps below this common level before receiving any resources. Allocating resources to them would yield less gain than providing further support to components with larger remaining gaps. These components consequently receive no resources in the optimal allocation. Proportional allocation generally does not achieve this balance, leaving scope to improve the overall gain through resource redistribution.

### B. *Optimal Individualized Allocation*

We consider a single student and omit the student index for clarity. Denote the positive post-ST mastery gap of component *j* by $g_j = 1 - k_{S,j}$. Given the receptivity $\alpha > 0$, and total AI learning budget $\tau_A > 0$, we seek an allocation that maximizes the average mastery increment:

$$\max_{\mathbf{w}_A} J(\mathbf{w}_A) = \frac{1}{d}\sum_{j=1}^{d} g_j\left[1 - \exp\left(-\alpha\tau_A w_{A,j}\right)\right], \quad (14)$$

subject to

$$w_{A,j} \geq 0, \quad \sum_{j=1}^{d} w_{A,j} = 1.$$

The objective is the average mastery increment across components, obtained from the update rule defined in Eq. (9). Since $\tau_A w_{A,j}$ is the actual learning budget directed to component *j*, the constraints ensure that the entire budget is distributed across components.

Let $\Delta k_j = k_{A,j} - k_{S,j}$ denote the learning increment of component *j*. Differentiating $\Delta k_j$ with respect to the allocation weight gives

$$\frac{\partial \Delta k_j}{\partial w_{A,j}} = \alpha\tau_A g_j \exp\left(-\alpha\tau_A w_{A,j}\right) = \alpha\tau_A\left(1 - k_{A,j}\right). \quad (15)$$

The marginal gain is thus proportional to the residual mastery gap $1 - k_{A,j}$, with the same coefficient $\alpha\tau_A$ for all components. Since all positively allocated components share the same multiplier associated with the normalization constraint, the KKT stationarity condition requires them to have equal marginal gains at the optimum. Components assigned zero resources can have marginal gains at zero allocation no greater than this common level. Denoting this common marginal level by *v*,

$$w_{A,j}^* > 0 \Rightarrow \left.\frac{\partial \Delta k_j}{\partial w_{A,j}}\right|_{w_{A,j}^*} = v, \quad (16a)$$

$$w_{A,j}^* = 0 \Rightarrow \left.\frac{\partial \Delta k_j}{\partial w_{A,j}}\right|_{w_{A,j}^*} \leq v. \quad (16b)$$

By Eq. (15), the common marginal-gain condition for positively allocated components gives

$$\alpha\tau_A\left(1 - k_{A,j}\right) = v,$$

and hence

$$\left(1 - k_{A,j}\right) = \frac{v}{\alpha\tau_A}.$$

Therefore, all positively allocated components have the same residual mastery gap. Denoting this common gap by

$$t = \frac{v}{\alpha\tau_A},$$

the KKT conditions can be written as

$$\begin{cases} g_j \exp\left(-\alpha\tau_A w_{A,j}^*\right) = t, & w_{A,j}^* > 0, \\ g_j \leq t, & w_{A,j}^* = 0. \end{cases} \quad (17)$$

The first condition shows that all supported components are allocated resources until their residual gaps reach the same level *t*. The second condition shows that a component whose initial post-ST gap is already no greater than *t* receives no resource.

Taking logarithms in the first condition and incorporating the zero-allocation condition gives the optimal weights as a function of *t*:

$$w_{A,j}^* = \frac{1}{\alpha\tau_A}\left[\ln\left(\frac{g_j}{t}\right)\right]_+, \quad [z]_+ = \max(z, 0). \quad (18)$$

Substituting Eq. (18) into the allocation constraint

$$\sum_{j=1}^{d} w_{A,j}^* = 1$$

yields

$$\sum_{j=1}^{d}\left[\ln\left(\frac{g_j}{t}\right)\right]_+ = \alpha\tau_A. \quad (19)$$

The left-hand side is continuous and strictly decreasing over $0 < t < \max_j g_j$, ranging from infinity to zero. Equation (19) therefore determines a unique threshold *t*, which can be obtained through one-dimensional numerical root finding. Substituting this threshold into Eq. (18) determines the optimal weights.

Components with $g_j > t$ receive resources until their remaining gaps reach *t*, whereas components with $g_j \leq t$ receive no resources. Thus, the resulting post-AI gaps satisfy

$$1 - k_{A,j}^* = \min\{g_j, t\}. \quad (20)$$

Finally, the objective has zero cross-component second derivatives, and

$$\frac{\partial^2 J}{\partial w_{A,j}^2} = -\frac{(\alpha\tau_A)^2}{d} g_j \exp\left(-\alpha\tau_A w_{A,j}\right) < 0. \quad (21)$$

Its Hessian is negative definite, making the objective strictly concave. Since the feasible set is convex, the allocation satisfying the KKT conditions is the unique global optimum.

### C. *Comparison with Proportional Allocation*

A controlled single-student experiment illustrates the difference between the two allocation rules. We consider *d*=10 components with the post-ST gap vector

$$\mathbf{g} = [0.90, 0.80, 0.65, 0.50, 0.35, 0.20, 0.12, 0.08, 0.04, 0.02].$$

The components are arranged in descending order of gap solely to facilitate comparison. The receptivity is set to $\alpha = 0.8$, and the total AI learning budget is $\tau_A = 2$. Both allocations use the same state, budget, and learning-response function.

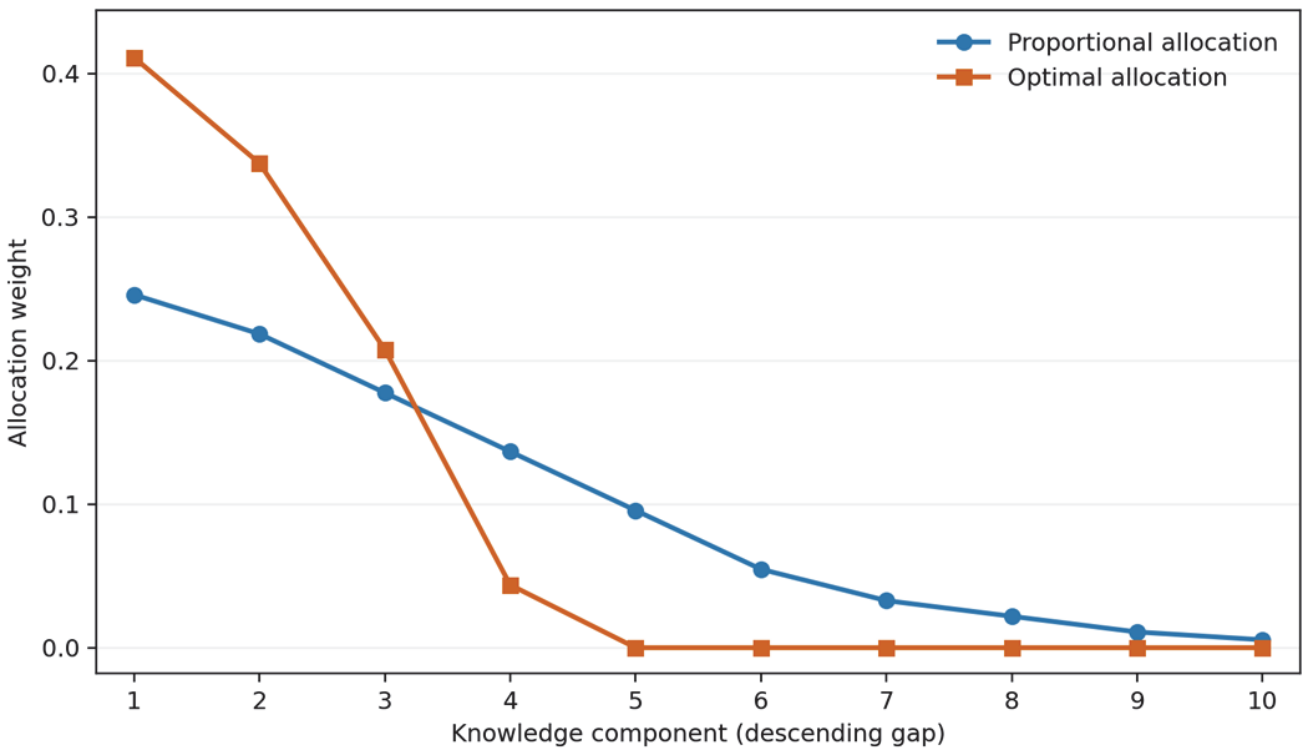

Fig. 6. Proportional and optimal AI allocation weights for ten knowledge components ordered by descending post-ST mastery gap.

In Fig. 6, proportional allocation assigns resources to all ten components, while the optimal allocation concentrates resources on the first four components, with weights approximately [0.4111, 0.3375, 0.2077, 0.0437], and assigns zero weight to components 5–10. The corresponding threshold is $t \approx 0.4662$. The average mastery increment is 0.0864 under proportional allocation and 0.0985 under optimal allocation, corresponding to an absolute increase of 0.0121 and a relative improvement of over 14%. This example demonstrates how concentrating a limited budget on components with higher marginal gains can improve the overall outcome, even though some components with positive gaps receive no additional support.

### D. *Implications for Resource Allocation*

The proportional rule provides a simple and intuitive baseline, whereas the optimal rule explicitly coordinates the use of a constrained budget. Unlike proportional weights, the optimal weights depend on the total budget, learner receptivity, and initial mastery gaps. As the budget increases, the threshold decreases and additional components may become eligible for support. With sufficiently large budgets, both proportional and optimal allocations drive all gaps toward zero, so their difference tends to vanish. Moreover, zero allocation indicates that, under the stated objective and budget, supporting that component yields less marginal gain than allocating the same resource elsewhere; it does not imply lower educational importance. Finally, the optimality result relies on the model assumptions of separable components, common receptivity across components, and an equally weighted average mastery objective.

## VII. Conclusion

This study develops a unified quantitative model of a continuous learning process comprising MOOCs, Smart Teaching, and AI-assisted learning. The three stages share a common exponential learning-response rule but differ in their resource-allocation mechanisms: predefined course-content emphasis, class-level adaptive allocation, and individualized support. This formulation makes the instructional process quantitatively analyzable, enabling stage-wise decomposition of learning gains, budget sensitivity analysis, examination of course–learner alignment, and optimization of individualized resource allocation.

Simulations across 100 independently generated classes demonstrate stable cumulative learning progression, while budget analysis shows diminishing returns from additional AI support. With the learner profile and learning mechanism fixed, the analysis shows a strong positive association between course–learner alignment in terms of cosine similarity and post-MOOC mastery level. Under a fixed AI budget, the optimal allocation equalizes the residual mastery gaps of supported components and assigns no resources to components already below that common level. A controlled comparison produces over 14% greater learning gain than proportional allocation. These results provide a quantitative basis for examining how course emphasis matches learner needs and how limited learning resources can be distributed more effectively.

Practical application requires connecting the model's variables to observable educational evidence and instructional decisions. Assessment data and cognitive diagnosis models [33] could provide estimates of component-level mastery, while the distribution of instructional time and activities could inform the course-content emphasis vector. Observed learning changes under recorded resource inputs could then be used to calibrate the response parameters. Model allocations could be translated into learning time or practice assignments and evaluated against existing allocation rules. This provides a feasible path from quantitative analysis to real-world teaching practice, although the educational effectiveness of such recommendations remains to be empirically validated.

Future work could incorporate prerequisite dependencies among knowledge components, changes in learner receptivity, and feedback through which learner states inform subsequent course design and resource allocation. Learner states could also distinguish dimensions such as knowledge, applied ability, and higher-order thinking, helping clarify the distinct roles of the three stages and why their integration is needed.

## Acknowledgment

The authors acknowledge the use of AI tools for language editing and manuscript preparation. All resulting text was reviewed by the authors, who take full responsibility for the content of this work.

## References

[1] A. Kaplan and M. Haenlein, "Higher education and the digital revolution: About MOOCs, SPOCs, social media, and the Cookie Monster," Business Horizons, vol. 59(4), pp. 441–450, 2016.

[2] M. Saini and N. Goel, "How smart are smart classrooms? A review of smart classroom technologies," ACM Computing Surveys, vol. 52(6), Article: 130, 2019.

[3] W. Holmes, M. Bialik, and C. Fadel, Artificial Intelligence in Education: Promises and Implications for Teaching and Learning. Boston, MA, USA: Center for Curriculum Redesign, 2019.

[4] I. Adeshola and A. Adepoju, "The opportunities and challenges of ChatGPT in education," Interactive Learning Environments, vol. 32(10), pp. 6159–6172, 2024.

[5] T. Zhang and B. Yuan, "Visualizing MOOC user behaviors: A case study on XuetangX," 17th International Conference on Intelligent Data Engineering and Automated Learning, pp. 89–98, 2016.

[6] G. Siemens, "Learning analytics: The emergence of a discipline," American Behavioral Scientist, vol. 57(10), pp. 1380–1400, 2013.

[7] S. Gill et al., "Transformative effects of ChatGPT on modern education: Emerging era of AI chatbots," Internet of Things and Cyber-Physical Systems, vol. 4, pp. 19–23, 2024.

[8] D. R. Sadler, "Formative assessment and the design of instructional systems," Instructional Science, vol. 18(2), pp. 119–144, 1989.

[9] J. Biggs, "From theory to practice: A cognitive systems approach," Higher Education Research & Development, vol. 12(1), pp. 73–85, 1993.

[10] L. Pan, C. Li, J. Li, and J. Tang, "Prerequisite relation learning for concepts in MOOCs," 55th Annual Meeting of the Association for Computational Linguistics, pp. 1447–1456, 2017.

[11] B. J. Zimmerman, "Becoming a self-regulated learner: An overview," Theory Into Practice, vol. 41(2), pp. 64–70, 2002.

[12] R. Liu and K. R. Koedinger, "Closing the loop: Automated data-driven cognitive model discoveries lead to improved instruction and learning gains," Journal of Educational Data Mining, vol. 9(1), pp. 25–41, 2017.

[13] I. Molenaar, "Towards hybrid human-AI learning technologies," European Journal of Education, vol. 57(4), pp. 632–645, 2022.

[14] K. Wang and C. Zhu, "MOOC-based flipped learning in higher education: students' participation, experience and learning performance," International Journal of Educational Technology in Higher Education, vol. 16, Article: 33, 2019.

[15] S. Pertuz, O. Reyes, E. Cristobal, R. Meier, and M. Castro, "MOOC-based flipped classroom for on-campus teaching in undergraduate engineering courses, " IEEE Transactions on Education, vol. 66(5), pp. 468–478, 2023.

[16] J. Kim, P. J. Guo, D. T. Seaton, P. Mitros, K. Z. Gajos, and R. C. Miller, "Understanding in-video dropouts and interaction peaks in online lecture videos," 1st ACM Conference on Learning @ Scale, pp. 31–40, 2014.

[17] A. F. Wise and Y. Jung, "Teaching with analytics: Towards a situated model of instructional decision-making," Journal of Learning Analytics, vol. 6(2), pp. 53–69, 2019.

[18] A. Alfoudari, C. Durugbo, and F. Aldhmour, "Understanding socio-technological challenges of smart classrooms using a systematic review," Computers & Education, vol. 173, Article: 104282, 2021.

[19] L. Yan et al., "Practical and ethical challenges of large language models in education: A systematic scoping review," British Journal of Educational Technology, vol. 55(1), pp. 90–112, 2024.

[20] E. Kasneci et al., "ChatGPT for good? On opportunities and challenges of large language models for education," Learning and Individual Differences, vol. 103, Article: 102274, 2023.

[21] M. V. Yudelson, K. R. Koedinger, and G. J. Gordon, "Individualized Bayesian knowledge tracing models," 16th International Conference on Artificial Intelligence in Education, pp. 171–180, 2013.

[22] C. Piech et al., "Deep knowledge tracing," 29th International Conference on Neural Information Processing Systems, pp. 505–513, 2015.

[23] A. N. Rafferty, E. Brunskill, T. L. Griffiths, and P. Shafto, "Faster teaching via POMDP planning," Cognitive Science, vol. 40(6), pp. 1290–1332, 2016.

[24] M. Schneider and F. Preckel, "Variables associated with achievement in higher education: A systematic review of meta-analyses," Psychological Bulletin, vol. 143(6), pp. 565–600, 2017.

[25] E. Bartl and R. Bělohlávek, "Knowledge spaces with graded knowledge states," Information Sciences, vol. 181(8), pp. 1426–1439, 2011.

[26] M. Richardson, C. Abraham, and R. Bond, "Psychological correlates of university students' academic performance: A systematic review and meta-analysis," Psychological Bulletin, vol. 138(2), pp. 353–387, 2012.

[27] A. Heathcote, S. Brown, and D. J. K. Mewhort, "The power law repealed: The case for an exponential law of practice," Psychonomic Bulletin & Review, vol. 7(2), pp. 185–207, 2000.

[28] A. T. Corbett and J. R. Anderson, "Knowledge tracing: Modeling the acquisition of procedural knowledge," User Modeling and User-Adapted Interaction, vol. 4, pp. 253–278, 1994.

[29] J. Lee and H. Choi, "Rethinking the flipped learning pre-class: Its influence on the success of flipped learning and related factors," British Journal of Educational Technology, vol. 50(2), pp. 934–945, 2019.

[30] I. Molenaar and C. Knoop-van Campen, "How teachers make dashboard information actionable," IEEE Transactions on Learning Technologies, vol. 12(3), pp. 347–355, 2019.

[31] G. Kestin, K. Miller, A. Klales, T. Milbourne, and G. Ponti, "AI tutoring outperforms in-class active learning: An RCT introducing a novel research-based design in an authentic educational setting," Scientific Reports, vol. 15, Article: 17458, 2025.

[32] P. Brusilovsky, "Adaptive navigation support in educational hypermedia: The role of student knowledge level and the case for meta-adaptation," British Journal of Educational Technology, vol. 34(4), pp. 487–497, 2003.

[33] J. de la Torre, "The generalized DINA model framework," Psychometrika, vol. 76(2), pp. 179–199, 2011.